# Geometrical Optics via electromagnetic wave quantization

John P. Hernandez
Department of Physics and Astronomy,
University of North Carolina, Chapel Hill

The basic laws of geometrical optics can be deduced from energy-momentum conservation for electromagnetic waves, without other wave concepts. However, the concept of quanta is required; it arises naturally, hence such a hypothesis could have arisen earlier than it did historically. Measurements to determine the angles of reflection and refraction demand that each incident quantum be either reflected or refracted; such a separation is central to the experimental results.

INTRODUCTION

Maxwell's great achievement (briefly summarized, in this paragraph, for what follows): culminated in the derivation that electromagnetic waves are transverse and the square of their speed equals the reciprocal product of the permeability and permittivity of the medium ($\mu\varepsilon$). Hence, in a linear, homogeneous, isotropic, and non-absorbing medium, the speed of light is the vacuum value (c) divided by the index of refraction ($n = [\mu\varepsilon/\mu_o\varepsilon_o]^{(1/2)}$). Further, the ratio of the electric field amplitude to that of the magnetic field is also the speed of light. Such waves transport energy and linear momentum, along the "rays" (lines perpendicular to the constant phase surfaces). The relation between energy and momentum follows from the power flux due to the wave. The power flux (*i.e.* the magnitude of the Poynting vector), divided by the speed of light, gives the radiation force exerted over the same area, perpendicular to the rays. Thus, from the concept that force is the time rate of change of momentum, the ratio of energy density (u) and linear momentum density (p) is $u/p = (c/n)$. All this is discussed in a first course on electromagnetism [1].

In an elementary treatment of electromagnetism, the next topic to be treated, following the material summarized above, is geometrical optics, an application in which all dimensions are to be large compared to the wavelength and times also large compared to the inverse frequency. It seems desirable to immediately show the connection between geometrical optics and the previously studied energy and momentum in electromagnetism. It is well known that such an approach reverses the historical sequence in which physical optics, the fully wave-based approach, was used for this topic; and also, that physical optics is indeed required to calculate the fraction of the energy reflected, interference, and diffraction phenomena, though it is unable to explain the photo-electric effect. The present approach, an alternative to one using Huygens' Principle or equivalent, offers an important advantage in an elementary treatment: exploring the direct connection between energy and momentum of electromagnetic waves and geometrical optics – yet, to my knowledge, this argument is not routinely given. The fact that there is something missing and required, quantization, may be viewed as a further advantage, in that the treatment suggests a hypothetical route to its early discovery.

GEOMETRICAL OPTICS

To define the problem, geometrical optics considers a plane wave (really a macroscopically narrow beam), in a medium (of index $n_1$), which has certain total energy ($U_I$) and momentum ($P_I = U_I /(c/n_1)$) incident, for simplicity, on an arbitrary area of a single, flat, interface during some arbitrary time interval. The momentum is incident on the interface along rays whose angle to the surface normal is $\theta_1$. The interface separates the first medium from a second one, of index $n_2$. The reflected energy, from the same area and in the same time interval, can be labeled as $U_R$ and the reflected momentum ($U_R /(c/n_1)$) departs the interface along rays at an angle $\theta_R$ from the surface normal. Finally, the energy transmitted, through the area in the time interval, is labeled as $U_T$; its associated momentum ($U_T /(c/n_2)$) also departs from the interface, but now into the second medium, along rays at an angle $\theta_2$ from the surface normal. As no energy is absorbed by the surface or the media and, if the interface is externally held stationary against the incoming radiation pressure, no external work is done, energy conservation implies $U_I = U_R + U_T$. The independent variables $U_I$ and $R \equiv (U_R / U_I)$ are not usually specified, or required, in geometrical optics; thus, there is no need to discuss the polarization of the incident wave, relative to the interface.

The incident wave exerts pressure on the interface area which, for a static situation, is balanced by a single external force, one perpendicular to the interface. Thus, linear momentum parallel to the interfacial plane must be conserved. Note that such momentum conservation, perpendicular to the plane of incidence (that containing the incident and surface-normal directions), demands that, with no incident momentum in that direction, any reflected and transmitted momenta must cancel; it is not required that each be zero (however, the reflection symmetry across the plane of incidence adequately excludes such momenta). Further, in the plane of incidence, momentum conservation parallel to the interface demands:

$$[U_I /(c/n_1)] \sin \theta_1 = [U_R /(c/n_1)] \sin \theta_R + [U_T /(c/n_2)] \sin \theta_2 . \qquad (1)$$

Equivalently, using energy conservation, the above can be rewritten as:

$$[U_R /(c/n_1)] \sin \theta_1 + [U_T /(c/n_1)] \sin \theta_1 = [U_R /(c/n_1)] \sin \theta_R + [U_T /(c/n_2)] \sin \theta_2 . \qquad (2)$$

Collecting terms and dividing by $U_I$ yields:

$$R\, n_1 (\sin \theta_1 - \sin \theta_R) = (1-R)(n_2 \sin \theta_2 - n_1 \sin \theta_1). \qquad (3)$$

Apparently, this is the end of the discussion based on energy-momentum conservation and it is not quite enough to account for the experimental data.

The experimental bases of geometrical optics are the laws of reflection and refraction. They are valid for coherent or incoherent incident waves and relate the angles and the indices of refraction. The familiar experimental laws consist of the following: confinement of all rays to the plane of incidence, specular reflection: $\theta_1 = \theta_R$, and Snell's law: $n_2 \sin \theta_2 = n_1 \sin \theta_1$. These experimental laws can be immediately identified as consistent, but not identical, with conservation of energy and of momentum, parallel to the interface, from the argument given above which resulted in equations (1-3). To obtain an identity, there is an additional requirement:

equations (1-3) must be satisfied independently of R. If this were not the case, the equations would also admit solutions ($\theta_R$ and $\theta_2$ in terms of R and $\theta_1$) in conflict with experiment, unless R=0 or 1. For the very specific cases, R=1 (total internal reflection) or R=0 (incidence at Brewster's angle with the incident polarization in the plane of incidence), there is only one outgoing term and the conservation equations have no solution other than the experimental one.

Given the experimental observations, equation (2) must be taken to imply that those fractions of the incident wave which are in fact reflected and transmitted are measured to conserve energy and momentum parallel to the interface, *separately*. Confinement of the reflected and refracted rays to the plane of incidence can also be identified as a consequence of the separate conservation of momentum parallel to the interface (there are no incident momenta perpendicular to the plane of incidence), in addition to the symmetry argument. It will be shown later that the momentum change perpendicular to the interface can also be interpreted to separate. Classically one must be careful with the point of view noted for equation (2), a division of the incident energy-momentum into a part which will be reflected and one which will be transmitted is conceivable but *cannot* be associated with any property of the incident wave, the physics must be elsewhere. The only unmentioned classical physics, the polarization, cannot be made to carry the burden of such a division; even for a unique polarization, in general, the wave is not fully reflected or refracted. However, the separation of the outgoing waves is clear, since the directions of propagation are *experimentally measured, separately*. There is no hint, in (1-3), that experimental geometrical optics should have solutions which are independent of R. Since energy-momentum conservation parallel to the interface does not yield the experimental results unless there is R independence, a physical explanation is required.

The simplest physical hypothesis consistent with experiment could have been suggested much earlier in history than was actually the case and is now known to be correct: the energy and momentum in electromagnetic waves are quantized and, in geometrical optics, *each* incident quantum is *either* reflected or refracted (for each quantum R=0 or 1, on measurements; these alternatives are probabilistic, thus avoiding the need for preselection in the incident wave). It then follows that if $U_I$ is the energy of such a quantum, only the first or the second term of the right side of (1) can be non-vanishing. The single-quantum argument is finally extended to a $U_I$ which contains an arbitrary number of identical quanta. The experimental laws are then a consequence of the conservation laws, regardless of R for the ensemble (though the macroscopic reflectivity of the interface is not available from the conservation laws, it is not required to obtain the laws of reflection and refraction). Finally, another result is implicit in the above: energy conservation, in the experiments of interest, requires that all quanta (incident, reflected, and refracted) have the same energy; this energy must therefore be a function of the frequency of the wave, the only factor identical in both media.

As an aside from the topic under consideration, it is noted that the external force, balancing the radiation pressure, integrated over the time interval, provides for the change in momentum perpendicular to the interface (final minus initial):

$$(\Delta p)_\perp = \{[U_R/(c/n_1)]\cos\theta_R - [U_T/(c/n_2)]\cos\theta_2\} + [U_I/(c/n_1)]\cos\theta_1$$
$$= U_I <[R/(c/n_1)][\cos\theta_1 + \cos\theta_R] + (1-R)\{[1/(c/n_1)]\cos\theta_1 - [1/(c/n_2)]\cos\theta_2\}>. \quad (4)$$

The effect is difficult to measure, it does depend on $U_I$ and R, and it is not included in the study of geometrical optics. The last line of (4) can be seen to be interpretable with the two separate processes: the term with R is the momentum change of any reflected quanta while the one with (1-R) is that due to the transmitted ones.

Also, it may be useful to reassure the casual reader: the above argument is compatible with now-known quantum concepts in such experiments [2]. In order to experimentally determine the directions of the reflected and refracted rays, *measurements* are required. Given that the observations are independent of the intensity of the incident radiation, such measurements guarantee that each photon *is* detected as entirely reflected or refracted, since no partial quanta are found in the measurements, in agreement with quantum mechanics. In physical optics experiments, for example with thin films, each photon *can* be though of as partially reflected and refracted to give interference effects, but a *measurement,* to determine the refraction path in the film, will destroy the interference since it is well known that number and phase operators do not commute (such an effect is well known in other contexts). Also, momentum uncertainties, say introduced by slits, *do* yield diffraction effects but are unobservable for the large dimensions, compared to the wavelength, considered in geometrical optics. Energy conservation is also compatible with the large measurement times, in such experiments. The present argument is, of course, based on the now-known particle picture of electromagnetism, whereas physical optics is based on the wave picture. Electromagnetism, as a particular case of natural phenomena, is subject to quantum wave-particle duality (with no intrinsic conflict); but different experiments show different aspects of this duality.

CONCLUSION

Newton gave a corpuscular argument, of the present type, to account for the laws of geometrical optics. His explanation suffered from the problem that he had the incorrect energy-momentum relation for electromagnetism. The correct energy-momentum argument given here is based on knowledge available to Maxwell, indeed obtained by him and his immediate successors, up to the missing quantization. Given that, during Maxwell's time, the wave description gave all experimental results discussed here and more, it would have required a great leap to put such knowledge aside and try to account for the experiments, in geometrical optics, on the basis of energy-momentum conservation exclusively. Historically, the time to consider the duality in the two pictures had not yet come. The concept of photons had to wait for a *conflict* between known theory and experiment: the photo-electric effect. It is curious that Einstein, who did have all the concepts available in his miraculous year (1905), also failed, I believe, to make the connection to geometrical optics. To summarize the present argument, the laws of geometrical optics can be seen to be identical to energy-momentum conservation for photons.